\documentclass[a4paper, 12pt]{article}

\usepackage[latin1]{inputenc}
\usepackage[english]{babel}

\usepackage{graphicx}
\usepackage{amsmath}
\usepackage{latexsym}
\usepackage{amssymb} 
\usepackage{epsfig}
\usepackage{cancel}
\usepackage{exscale}

\newcommand{\Si}{\scriptstyle}

\begin{document}

\begin{titlepage}
\vspace{1.5cm}
\begin{center}
{\LARGE \bf A Lagrangean formalism for\\Hermitean matrix models } \\  
\vspace{2cm}

{ \large R. Flume$ ^1 $, J. Grossehelweg$ ^2
$, A. Klitz$ ^1 $}
\\ 
\vspace{1.5cm}
$ ^1 $ Physikalisches Institut, Universit\"at Bonn,\\
Nu{\ss}allee 12, 53115 Bonn, Germany\\ 
\vspace{0.2cm}
$ ^2 $ DESY Theory Group, DESY Hamburg,\\
 Notkestr. 85, 22603 Hamburg, Germany\\

\vspace{2cm}
\begin{abstract}
 Eynard's formulation of Hermitean 1-matrix models in terms of intrinsic
 quantities of an associated hyperelliptic Riemann surface is rephrased as a
 Lagrangean field theory of a scalar particle propagating on the hyperelliptic
 surface with multiple self-interactions and particle-source-interactions. Both
 types of interaction take place at the branch points of the hyperelliptic
 surface.
\end{abstract}

\end{center}
\end{titlepage}

\newpage

\section{Introduction}

The conceptual and practical understanding of some random matrix models has
been advanced significantly by the seminal work of Eynard
\cite{4}. This author introduced an elegant new approach to the topological
expansion of Hermitean 1-matrix models based on the intrinsic notions of a
hyperelliptic Riemann surface which is---as spectral surface---attached to the
matrix model. The new approach has been further elucidated in \cite{5} and
applied to 2-matrix models in \cite{6} and \cite{7}. It has moreover been
shown in \cite{8} that one may abstract from the framework of matrix models to
the construction of new invariants attached to more general Riemann surfaces
(mimicking the topological expansion of matrix models). This brings into focus
\cite{8} new subjects as Kontsevich's matrix model for intersection numbers
\cite{9}, topological string theory \cite{10}, \cite{11}, \cite{12} and
\cite{13}, as well as further applications to mathematical subjects, as the
recursive determination of Weil-Petersson volumes \cite{14}, \cite{15} and a
conjectural matrix model representation of Hurwitz numbers \cite{16}.\\ \\
The purpose of this note is to show that Eynard's elegant construction can be
deciphered into a non-elegant effective Lagrangean formalism. We will
concentrate here on Hermitean 1-matrix models---but we hope that our methods
will also apply in some of the instances mentioned above. The Lagrange to be
constructed will be that of a scalar field propagating on a hyperelliptic
Riemann surface (i.e. the Riemann surface given as the spectral curve
associated with the matrix model). The corresponding 'vacuum graphs' of the
Gell-Mann-Low expansion and the graphs for the Green's functions will
be shown to represent the free energy and the correlations of resolvent
operators resp. of the matrix model. The self-interaction of the
scalar field and its interaction with background sources will be given as
formal infinite power series, which receive 'short distance' (see below)
corrections throughout the loop expansion of the effective theory. This latter
expansion will be identified with the topological expansion of the matrix
model. \\The unpublished work \cite{19} has been fused into the present article.\\ \\ 
The plan of the paper is as follows: In section 2 we collect some technical
material concerning Hermitean 1-matrix models taken from the literature
\cite{1} - \cite{4}, \cite{17}.
(Experts may skip this section.) Section 3 is devoted to the construction of
the effective Lagrange in tree graph approximation which is identified with
the leading large N order for the matrix model. In section 4 we discuss the
non-leading terms of the matrix model topological expansion, i.e. the loop corrections in Lagrangean parlance. We end in section 5 with some conclusions.

\section{Hermitean 1-matrix models}

The partition function of Hermitean 1-matrix models (\cite{1}, \cite{2}) is
given by a (potentially formal) integral over Hermitean matrices 
\begin{eqnarray}
\label{eq: partition sum}
 Z_N (t) = \int  e^{-N \; {\rm tr} \; V(M) } \; dM \;,
\end{eqnarray}
\begin{displaymath}
 dM = \underset{i}{\prod} \; dM_{ii}  \;\; \underset{i<j}{\prod} \; Re
\ dM_{ij} \;\; \underset{i<j}{\prod} \; Im \  dM_{ij} \; ; 
\end{displaymath}
$ M_{ii} $ are real variables; the
potential $ V(M) $ is given by $ V(M) = \sum_{n \geq 1} t_n M^n $, with $ t_n
$ denoting some coupling constants. Ensuing averages are introduced as usual:
\begin{equation}
\langle f(M) \rangle  = \frac{1}{Z} \int f(M) \; e^{-N \; {\rm tr} \; V(M) }
 \; dM  .
\end{equation}
The free energy $ \mathcal{F} $ related to the partition function by $ Z_N
(t) = e^\mathcal{F} $ can formally be expanded in powers of $ \frac{1}{N} $ (the
so-called topological expansion)
\begin{equation} 
 \mathcal{F}= \sum_{h=0}^\infty N^{2-2h} \; \mathcal{F}^{(h)}.  
\end{equation}
Following well established paths \cite{3}  we introduce a derivative called 
the 'loop operator' 
\begin{equation} 
\label{eq: loop operator}
 \frac{ \partial }{ \partial V (z) } = - \sum_{j=1}^\infty \frac{1}{z^{j+1}} 
\frac{\partial}{\partial t_j}, 
\end{equation} 
z denoting here a complex number. Connected correlators of the resolvent 
\begin{displaymath}  
W(z) = \frac{1}{N} tr \frac{1}{z-M}  
\end{displaymath}
are obtained by multiple application of the loop operator to the free 
energy
\begin{equation}   
W_n (z_1, \ldots ,z_n) = N^{2n-2} \langle W(z_1) \ldots W(z_n) \rangle_{\rm conn} = \frac{1}{N^2}
\frac{ \partial }{\partial V (z_1) } \cdots \frac{ \partial }{\partial V (z_n)
  } \mathcal{F} .
\end{equation}
The subscript 'conn' refers to the connectedness of the correlator.\\
The correlation function $ W_n $ inherits from the free energy a topological
expansion
\begin{equation}
\label{eq: top.exp. of W}
 W_n (z_1, \ldots, z_n) = \sum_{h=0}^\infty N^{-2h}  W_n^{(h)} (z_1, \ldots, z_n). 
\end{equation} 
It turned out that the most efficient tool for the determination of the
partition function (the free energy resp.) and of the correlators
of resolvents are the so-called loop identities, that is, Ward identities for
invariance of the defining integrals under reparametrisations of the matrix
model integration variables of the type
\begin{displaymath}
 \delta M = \varepsilon \frac{1}{z-M}  
\end{displaymath}
with $ \varepsilon $ denoting an infinitesimal real parameter and $ z $ an
arbitrary  complex number. The invariance of the partition function under this
variation leads to the identity
\begin{equation}
\label{eq: loop eq. of order 0}
(W_1(z))^2 + \frac{1}{N^2} W_2 (z,z) = \frac{ 1 }{ N } \langle    {\rm tr}
 \frac{   V^\prime (M)     }{z-M} 
 \rangle . 
\end{equation}
The l.h.s. of the latter equation derives from the variation of the measure in
eq. (\ref{eq: partition sum}) under the variation (\ref{eq: top.exp. of W}) whereas the r.h.s. derives from the variation
of the integrand $ \exp(-N \; {\rm tr} \, V(M) ) $. Applying to eq. (\ref{eq: loop
  eq. of order 0}) the loop
operator---or equivalently, starting from a more involved variation
substituting instead of eq. (\ref{eq: top.exp. of W}) 
\begin{displaymath}
 \delta M = \varepsilon \frac{1}{z_1 - M} \prod_{j > 1} {\rm tr} 
\frac{1}{z_j-M}  
\end{displaymath}
one can deduce further identities for multipoint functions. Eynard \cite{4}
uses those identities to establish for the correlation functions and finally
also for the higher order corrections to the free energy a recursive
procedure giving rise to a trivalent graphical representation (cf. \cite{5} in the
latter context).\\
Invariance under the unitary group $ U(N) $ may be exploited to reduce all $
N^2 $-dimensional integrals above to $ N $-dimensional integrals over
eigenvalues of the Hermitean matrices. The partition function, eq. (\ref{eq:
  partition sum}), in
particular becomes then \cite{1} 
\begin{equation}
\label{eq: partition
  function in eigenvalue notation}
Z_N(t) = \int \prod_{i=1}^N d\lambda_i \prod_{i<j} (\lambda_i - \lambda_j)^2 
\prod_i e^{-N V(\lambda_i)} , 
\end{equation}
where we have skipped---for convenience---on the r.h.s. of (\ref{eq: partition
  function in eigenvalue notation}) the volume
factor of
the unitary group. We assume (as \cite{1}-\cite{4}) that
the eigenvalues are, for large values of $ N $, concentrated on a finite
number of compact intervals on the real axis, say
\begin{displaymath}
\bigcup_{l=1}^s \; A_l \mbox{ with } A_l =  [a_{2l-1}, a_{2l}], \ \ \ a_1 < a_2 < \ldots < a_{2s}    
\end{displaymath}
where the separate intervals $ [a_{2l-1}, a_{2l}] $ are supposed to be 
spread in the neighbourhood of diverse minima of the potential $ V(\lambda)
$. The spectral density $ \rho_N (\lambda) $, concentrated on the above
intervals, becomes a continuous function $ \rho (\lambda) $ in the large $ N $
limit. The expectation value of the resolvent reads in terms of this
distribution as 
\begin{equation}
 W_1 (p) = \int d\lambda \frac{\rho(\lambda)}{p-\lambda} 
\end{equation}
and the loop equation (\ref{eq: loop eq. of order 0}) becomes 
\begin{equation}
\label{eq: loop eq. in integral rep.} 
 \oint\limits_{\mathcal{C}} \frac{d\omega}{2 \pi i} \frac{V^\prime(\omega)}{p-\omega} W_1
(\omega) = (W_1 (p))^2 + \frac{1}{N^2} W_2 (p,p) .
\end{equation}
The contour $ \mathcal{C} $ of the integral on the l.h.s. is supposed to
encircle in the complex plane the support of the spectral measure, but not the
point $ p $. The
second term on the r.h.s. of (\ref{eq: loop eq. in integral rep.}) drops out in leading order of the large $ N
$ expansion and one is left with the quadratic equation for the leading part $
W_1^{(0)} $,
\begin{displaymath}
 (W_1^{(0)} (p))^2 = \oint\limits_{\mathcal{C}} \frac{d\omega}{2 \pi i} 
\frac{V^\prime (\omega)}{p - \omega} W_1^{(0)} (\omega)  = \int d \omega
\frac{ V^\prime ( \omega ) - V^\prime  (p) + V^\prime (p) }{ p- \omega } \rho (\omega) 
\end{displaymath}
\begin{equation}
\label{eq: loop eq. with potential}
 = V^\prime (p) \; W_1^{(0)}(p) -  \int d\omega  \frac{ V^\prime (p)  - V^\prime( \omega ) }{p-\omega} \rho( \omega )  =
 V^\prime (p) \; W_1^{(0)}(p) + f(p),   
\end{equation}  
where $ f(p) $ denotes a polynomial of a degree one less than the degree of $ 
V^\prime (\omega) $. $ W_1^{(0)} (p) $ is given as solution of the preceding
quadratic equation, 
\begin{equation}
 W_1^{(0)} (p) = \frac{1}{2} V^\prime (p) - \frac{1}{2} \sqrt{(V^\prime (p))^2 + 4
  f(p) }, 
\end{equation}
the sign of the square root term being dictated by the stipulated asymptotic
behaviour $ W_1^{(0)} (p) \sim \frac{1}{p} $, $ p \rightarrow \infty $.
The expectation value of the resolvent is in this way related to a
hyperelliptic Riemann surface given by the equation 
\begin{equation} 
 y^2 = (V^\prime (\lambda) )^2 + 4 f (\lambda) . 
\end{equation}
One is free to choose $ f $ s.t. the polynomial on the r.h.s. has a finite
even number, say $ 2s $, of
simple zeros in accordance with the above assumption on the support of the 
spectral density---besides an appropriate number of double zeros, s.t. $ y $ reads
as 
\begin{equation}
\label{eq: y}
 y = M(\lambda) \tilde{y} ,
\end{equation}
\begin{equation}
\label{eq: reduced y}
\tilde{y}^2 = \prod_{i=1}^{2s} (\lambda - a_i), 
\end{equation}
where $ \tilde{y} $ now represents a reduced hyperelliptic surface 
\cite{5} of the same genus $ g=s-1 $ as $ y $. The polynomial $ M(\lambda) $
 can be seen by inspection of the loop equation (\ref{eq: loop eq. with potential}) to be given by a Cauchy  
contour integral (cf. \cite{3})  
\begin{equation} 
 M(\lambda) = \oint\limits_\mathcal{C_\infty} \frac{d \omega}{2 \pi i} \frac{1 }{\omega -\lambda } 
\frac{V^\prime (\omega) }{\sqrt{\prod_{i=1}^{2s} (\omega - a_i ) }}   
\end{equation}
where $ \mathcal{C_\infty} $ denotes a contour encircling all branch points $ 
a_i \ , i=1, \ldots , 2s $ and $ \lambda $. $ W_1^{(0)} (\lambda) $ is
found similarly
\begin{displaymath} 
W_1^{(0)}(\lambda) = \frac{1}{2} \oint\limits_\mathcal{C} \frac{d\omega}{2 \pi i}  
\frac{V^\prime (\omega) }{\lambda-\omega} \sqrt{\frac{\prod_{i=1}^{2s}  
    (\lambda-a_i)}{\prod_{j=1}^{2s} (\omega-a_j)}}   
\end{displaymath} 
where $ \mathcal{C} $ now encircles the branch points but not $\lambda$. \\ 
To fix the positions of the branch points one has to impose particular boundary
conditions. Following \cite{4}, \cite{17} we choose those to be 
given by 'filling fractions' 
\begin{displaymath}
 \varepsilon_l = \frac{1}{2} \oint\limits_{A_l} \frac{d \lambda }{2 \pi i} y( \lambda )
\end{displaymath}
which are the
relative weights of the various spectral intervals. The insertion of the 
topological expansion (\ref{eq: top.exp. of W}) into the loop equation (\ref{eq: loop eq. in integral rep.}) gives rise to 
\begin{equation}
\label{eq: operator K}
 ( \hat{\mathcal{K}} - 2 W_1^{(0)} (p) ) W_1^{(h)} (p) = \sum_{h^\prime = 1 
}^{h-1} W_1^{(h^\prime)} (p) W_1^{(h-h^\prime)} (p) + W_2^{(h-1)} (p,p) ,
\end{equation}
\begin{displaymath}
 \hat{\mathcal{K}} f(p) =  \oint\limits_{\mathcal{C}} \frac{d\lambda}{2 \pi i} 
\frac{V^\prime (\lambda) }{p-\lambda} f(\lambda)  .
\end{displaymath}
The solution of (\ref{eq: operator K}) under the condition  
\begin{displaymath}
 \oint\limits_{A_j} dx_1 W_2 (x_1,x_2) = 0 \ \ \ \ \ \ \ \forall j 
\end{displaymath}
(stipulating that the filling
fractions are to be chosen independently of the form of the interaction 
potential  $ V(t) $) is found, \cite{4}, to be 
\begin{equation}
\label{eq: loop eq. to order
h}
 W_1^{(h)} (p) = \sum_{i=1}^{2s}  \underset{ x \rightarrow a_i }{\rm Res}
 \frac{ dS(p,x) }{ dp } \frac{1}{y(x)} 
\left( \sum_{m=1}^{h-1} W_1^{(h-m)} (x)  W_1^{(m)} (x) + W_2^{(h-1)} (x,x) \right) 
\end{equation}
with $ dS(p,x) $ denoting a meromorphic one form with respect to the variable
$ p $ and a multivalued meromorphic function in $ x $ on the reduced
Riemann surface (\ref{eq: reduced y}). $ dS $ is characterized uniquely by the following properties:
\begin{eqnarray*}  
dS(x,x^\prime ) &\underset{ x \rightarrow x^\prime }{ = }& \frac{dx}{x-x^\prime}
+ \mbox{ finite}, \\
 dS(x,x^\prime ) &\underset{ x \rightarrow \overline{x^\prime} }{ = }&  -
 \frac{dx}{x-x^\prime} + \mbox{ finite}, 
\end{eqnarray*}
($ \overline{x}$ denoting the hyperelliptic involution of $ x $) where the 
quoted singularities are the only ones on the reduced surface, and 
\begin{displaymath}
 \oint\limits_{A_j} dS(x,x^\prime) =0  \ \ \ \ \forall \; j =1,\ldots , s-1. 
\end{displaymath}
The explicit expression for $ dS(x,x^\prime) $ is
\begin{equation}
\label{eq: explicit expression for dS }
 dS(x, x^\prime) = \frac{\sqrt{\sigma(x^\prime)}}{\sqrt{\sigma(x)}}  \left(    
  \frac{1}{x-x^\prime} - \sum_{j=1}^{s-1} C_j (x^\prime) L_j (x) \right) dx 
\end{equation} 
\begin{displaymath} 
 \mbox{with\ \ } C_j (x^\prime) = \frac{1}{2 \pi i} \oint\limits_{A_j}  
\frac{dx}{\sqrt{ \sigma(x) }} \frac{1}{x-x^\prime} \mbox{ and } \sigma(x) =
\prod_{i=1}^{2s} (x-a_i) .  
\end{displaymath} 
$ L_j (x) $ denote polynomials which are constitutive parts of the  
 normalized holomorphic 1-forms of the reduced surface,
\begin{displaymath} 
 \omega_j (x) = \frac{L_j(x)}{\sqrt{ \sigma(x) }} dx 
\end{displaymath} 
\begin{displaymath}
 \oint\limits_{A_l}  \omega_j (x) = 2 \pi i \delta_{jl}
 \mbox{\ for 
  \ } l,j \in { 1,\ldots,  s-1}.  
\end{displaymath}  
One needs for the resolution of the recursion, eq. (\ref{eq: loop eq. to order
h}), in explicit form the 
two-point function $ W_2^{(0)} (p,q) $. This is found as solution of the loop equation for the two-point
function---which emerges from another application of the loop operator to both 
sides of eq. (\ref{eq: loop eq. with potential})---in terms of the 'Bergmann kernel' $ B(p,q) $
(cf. \cite{4}): 
\begin{equation}
\label{eq: two-point function}
 W_2^{(0)} (p,q) = \frac{B(p,q)}{dq \ dp} - \frac{1}{(p-q)^2}.  
\end{equation}
$B$ is a symmmetric bidifferential on the reduced surface with a unique second 
order pole at the coincidence point $ p=q $, $ B(p,q)= ( \frac{dp\;dq}{(p-q)^2} +
{\rm 
  finite } )
$ for $ p \rightarrow q $ and satisfying the normalization conditions
\begin{displaymath}
  \oint\limits_{ q \in A_j } B(p,q) = 0  , \ j=1, \ldots, s-1 .
\end{displaymath}
The Bergmann kernel is related to the above introduced differential $ dS $ as
follows:\\
\begin{equation}
\label{eq: relation between B and dS}
 B(p,q) = \frac{1}{2} \frac{\partial}{\partial q} \left( \frac{ dp }{ p-q }
  + dS (p,q) \right) dq . 
\end{equation}
The solution of the recursion relation (\ref{eq: loop eq. to order
h}) and its generalization
for multipoint correlation functions in terms of $ dS(p,q) $, $ B(p,q) $, $
y(q) $, i.e. in terms of intrinsic quantities of the hyperelliptic surfaces
(\ref{eq: y}) and (\ref{eq: reduced y}) is due to Eynard. For details of this formalism we refer to the
original work \cite{4} and \cite{5}, \cite{8}.

\section{ The planar approximation in terms of tree graphs of an effective
  Lagrange}

The determination of the multipoint correlators of resolvents in
leading order of $ \frac{1}{N} $ may either proceed by inspection of the
multipoint generalisation of the loop equation (\ref{eq: loop eq. with potential})---this approach has been
followed by Eynard \cite{4}---or may be achieved by repeated application of
the loop operator to the above noted result, eq. (\ref{eq: two-point function}), for the resolvent
two-point correlation function. That is, one applies the loop operator either to
the loop equation or to its solution. We will follow here the second route.\\
The variation of the Bergmann kernel, eq. (\ref{eq: relation between B and dS}), due to an infinitesimal change
of the branch points of the underlying hyperelliptic surface is given by one
of Rauch's variational formulas \cite{18}:
\begin{equation}
\label{eq: B variation}
\delta B (p,q) = \frac{1}{2} \sum_j B(p,[a_j]) B(q,[a_j]) \delta a_j .
\end{equation}

$ B (p, [a_j] ) $ governs the asymptotic behaviour
of $ B (p,q) $ in $ q $ near $ a_j $:
\begin{displaymath}
B (p, [a_j]) = \left. 2  \frac{ \sqrt{ q - a_j }   B (p,q) }{ dq } \right|_{q
  \rightarrow a_j } . 
\end{displaymath}

In the following we will use
\begin{displaymath}
\widetilde{B} (p,q) = \frac{ B(p,q) }{ dp \; dq } - \frac{1}{2}
\frac{1}{(p-q)^2}   
\end{displaymath}
and
\begin{displaymath}
\widetilde{B} (p, [a_j] ) =  2  \oint\limits_{a_j} \frac{dz}{2 \pi i} 
\frac{ \widetilde{B}  (p,z) }{ ( z - a_j )^{1/2} } .
\end{displaymath}

Rauch's formula is to be combined with an expression for $ \frac{\delta
  a_j}{\delta V(p)} $. The latter is found by first noting the relation, \cite{5}, \footnote{We remind in this context the identities $ y(q) = -2 W_1^{(0)} (q) + V^\prime
(q) $ and \\
$ \frac{ \delta }{ \delta V (p) } V^\prime (q)  = - \frac{1}{(q - p)^2} $, the latter
being a direct consequence of eq. (\ref{eq: loop operator}).}
\begin{equation}
\label{eq: y variation}
\frac{\delta y (q)}{\delta V (p)} = - 2 \widetilde{B} (p,q) 
\end{equation}
 
and to match this with the asymptotic behaviour $ y ( q ) = y ( [ a_i ] )
\sqrt{ q - a_i } + \mathcal{O} (( q - a_i )^{3/2}) $, which implies 
\begin{equation}
\label{eq: asymptotic y variation}
\left. \frac{ \delta y (q ) }{ \delta V (p) }  \right|_{q \rightarrow a_i} = - 
\frac{1}{2} y ( [ a_i ] )  \frac{ 1 }{ \sqrt{ q - a_i } } \frac{ \delta
  a_i }{ \delta V (p) } + \mathcal{O} ( \sqrt{ q - a_i } ) .
\end{equation}
Eqs. (\ref{eq: y variation}) and (\ref{eq: asymptotic y variation}) lead us to
\begin{equation}
\label{eq:
  branch point variation}
\frac{ \delta a_i }{ \delta V (p) } =  \frac{ 2 \widetilde{B} ( p , [ a_i ] ) }{ y ( [ 
  a_i ] ) }    
\end{equation}
which with eq. (\ref{eq: B variation}) gives rise to 
\begin{equation}
\label{eq: three-point function}
W_3^{(0)} (p_1, p_2, p_3) = \frac{ \delta }{ \delta V (p_3) } W_2^{(0)}
(p_1,p_2) = \sum_{i=1}^{2s} \frac{ \widetilde{B} (p_1, [a_i])  \widetilde{B}(p_2, [a_i]) \widetilde{B}(p_3, [a_i])
}{ y ([a_i]) } .
\end{equation}

\begin{figure}[h!] 
\begin{center}
\epsfig{figure=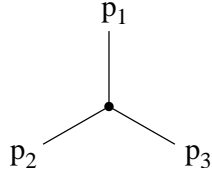} 
\caption{Three point function $ W_3^{(0)} $. The vertex factor is $ 1/y_{1,i}
  $. } 
\end{center}
\end{figure}

We will use the notations
\begin{eqnarray*}
\ \ \ \ \ \ \ \ \ \ B_i^f (p) & = & 2 \oint\limits_{a_i} \frac{dz}{ 2 \pi i } \frac{ \widetilde{B} ( p, z ) }{ ( z - a_i )^{f+1/2} }
\ \ \ \ \ \ \ \ \ \ \ \ \ \ \ \ \ \ \ \ \ \ \ \ \ \ \ \ \ \ \ \ \ \ \ \ \ \ \
\ \  {\rm(27a)}\\
B_{i,j}^{f,g} & = & 4 \oint\limits_{a_i} \frac{dz}{ 2 \pi i } \oint\limits_{a_j} \frac{
  dz^\prime }{ 2 \pi i } \frac{ \widetilde{B} (z,z^\prime) }{(z-a_i)^{f+1/2}
  (z-a_j)^{g+1/2} }  \ \ \ \ \ \ \ \ \ \ \ \ \ \ \ \,  {\rm(27b)}\\
y_{f,i} & = & \oint\limits_{a_i} \frac{dz}{ 2 \pi i} \frac{ y(z) }{ (z -
  a_i)^{f+1/2} } .\ \
\ \ \ \ \ \ \ \ \ \ \ \ \ \ \ \ \ \ \ \ \ \ \  \ \ \ \ \ \ \ \ \
\ \ \ \ \ \ \ \ \! \: {\rm(27c)}
\end{eqnarray*}
\addtocounter{equation}{1}
It will turn out that the collection of quantities (27a)-(27c) is complete in
the sense that the resolvent correlators and the free energy can be
expressed---as will be shown---to all orders in the topological expansion by those terms. One should note that (27b) is well defined also for $ i =
j $, since the order of the contours does not matter due to the fact that $
B(p,q) $ has as single singularity a second order pole with residuum
one. Straightforward calculations using eqs. (\ref{eq:
  branch point variation}) and (\ref{eq: three-point function}) lead to the relations
\begin{eqnarray}
\label{eq: y_fi variation}
\frac{ \delta y_{f,i}}{\delta V (q) } &=& \big( \Delta_1 (q) + \Delta_2 (q) 
\big) (y_{f,i}) 
\\
  &&\big( \Delta_1 (q) \big) (y_{f,i})= (2 f + 1) \frac{ y_{f+1,i}} {y_{1,i}}  
B_i^0 (q) \ \ \ \ \ \ \ \ \ \ \ \ \ \ \ \ \  {\rm (\ref{eq: y_fi variation}a)} 
\nonumber\\
&&\big( \Delta_2 (q) \big) (y_{f,i}) = - B_i^f (q) \ \ \ \ \ \ \ \ \ \ \ \ \ \
\ \ \ \ \ \ \ \ \ \ \ \ \ \ \ \ \ \ \ {\rm (\ref{eq:
    y_fi variation}b)} 
\nonumber\\ 
\label{eq: B_i^f variation} 
\frac{ \delta B_i^f (p) }{\delta V (q) } &=& \big( \Delta_3 (q)  +  \Delta_4
(q) \big) (B_i^f (p))
\\
&&\big( \Delta_3 (q) \big) (B_i^f (p))= (2 f + 1) \frac{ B_i^{f+1} (p) B_i^0
  (q) }{ y_{1,i} } \ \ \ 
\ \ \ \ \ \ \ {\rm (\ref{eq: B_i^f variation} 
a)}
\nonumber\\
&&\big( \Delta_4 (q) \big (B_i^f (p)) =  \sum_{j=1}^{2s} \frac{ B_{i,j}^{f,0}
  B_j^0 (p) B_j^0 (q) }{ y_{1,j} } \ \ \ \ \ \ \ \ \ \ \ \ \ \ \: {\rm (\ref{eq: B_i^f variation}
b)}
\nonumber\\
\label{eq: B_ij^fg variation} 
\frac{ \delta B_{i,j}^{f,g} }{ \delta V (q) }  &=& \big( \Delta_5 (q) + \Delta_6 (q)
+ \Delta_7 (q) \big) ( B_{i,j}^{f,g})
\\
&&\big( \Delta_5 (q) \big) ( B_{i,j}^{f,g}) = ( 2 f + 1 ) \frac{ B_{ i,j 
  }^{ f+1,g } B_i^0 (q) }{ y_{1,i} } \ \ \ \ \ \ \ \ \ \ \ \ \ \ \, {\rm (\ref{eq: B_ij^fg variation}a)}
\nonumber\\
&&\big( \Delta_6 (q) \big) ( B_{i,j}^{f,g}) = ( 2 g + 1 ) \frac{ B_{ i,j }^{f,g+1} 
  B_j^0 (q) }{ y_{1,j} }  \ \ \ \ \ \ \ \ \ \ \ \ \ \ \ \! \! \; {\rm (\ref{eq: B_ij^fg variation}b)}
\nonumber\\
&&\big( \Delta_7 (q) \big) ( B_{i,j}^{f,g}) = \sum_{k=1}^{2s} \frac{ B_{i,k}^{ f,0 } B_{ j,k }^{ g,0 }
  B_k^0 
  (q) }{ y_{1,k}} .\ \ \ \ \ \ \ \ \ \ \ \ \ \ \ \ \ \! \: {\rm (\ref{eq: B_ij^fg variation}c)}
\nonumber
\end{eqnarray}
The application of an operator $ \Delta_j (q)$, $ j = 1, \ldots, 7 $ to expressions not noted above gives zero. It can also
easily be verified that $ \frac{ \delta }{ \delta V (q) } $ and $ \frac{
  \delta }{ \delta V (p) } $ are for arbitrary arguments $ p $ and $ q $ 
commuting operators acting on the basis\linebreak[4] (27a-c) (as it should be 
indeed). The 4-point function is found by application of $
\frac{ \partial }{ \partial V (p) } $ (cf. eqs. (\ref{eq: y_fi variation}) and (\ref{eq: B_i^f variation})) to $ W_3^{(0)} $. One finds:
\begin{eqnarray}
&& \!\!\!\!\!\!\!\!\!\!\!\!\!\! W_4^{(0)} (p_1,\ldots,p_4) =\frac{ \delta }{ \delta V (p_4) } W_3^{(0)} 
(p_1,p_2,p_3) \hspace{3cm}
\nonumber\\
&=& \sum_{i,j=1}^{2s}  \left(  B_i^0 (p_1) B_i^0 (p_2) \left( \frac{ 1 }{ y_{1,i} } \right)
B_{i,j}^{0,0} \left( \frac{ 1 }{ y_{1,j} } \right)  B_j^0 (p_3) B_j^0 (p_4) \right.
\nonumber\\
 {} &+& B_i^0 (p_1) B_i^0 (p_3) \left( \frac{ 1 }{ y_{1,i} } \right)
B_{i,j}^{0,0} \left( \frac{ 1 }{ y_{1,j} } \right)  B_j^0 (p_2) B_j^0 (p_4)
\left. \right. 
\nonumber\\
 {} &+& \left. B_i^0 (p_1) B_i^0 (p_4) \left( \frac{ 1 }{ y_{1,i} } \right)
B_{i,j}^{0,0} \left( \frac{ 1 }{ y_{1,j} } \right)  B_j^0 (p_3) B_j^0 (p_2)
\right)
\nonumber\\
\label{eq: four-point function}
&+&\sum_{i=1}^{2s} \left( - 3 \frac{ y_{2,i} }{ y_{1,i}^3} \right)
\prod_{r=1}^4 B_i^0 (p_r)  +\sum_{r=1}^4 \sum_{i=1}^{2s}   \left( \frac{ 1 }{ y_{1,i}^2 } \right)
B_i^1 (p_r) \prod_{ \Si t = 1 \atop \Si t \neq r}^4 B_i^0 (p_t) .
\end{eqnarray}  
Eq. (\ref{eq: three-point function}) and (\ref{eq: four-point function}) designate the starting for an effective Lagrangean 
description of the resolvent operator correlation functions.
\begin{figure}[h] 
\begin{center}
\epsfig{figure=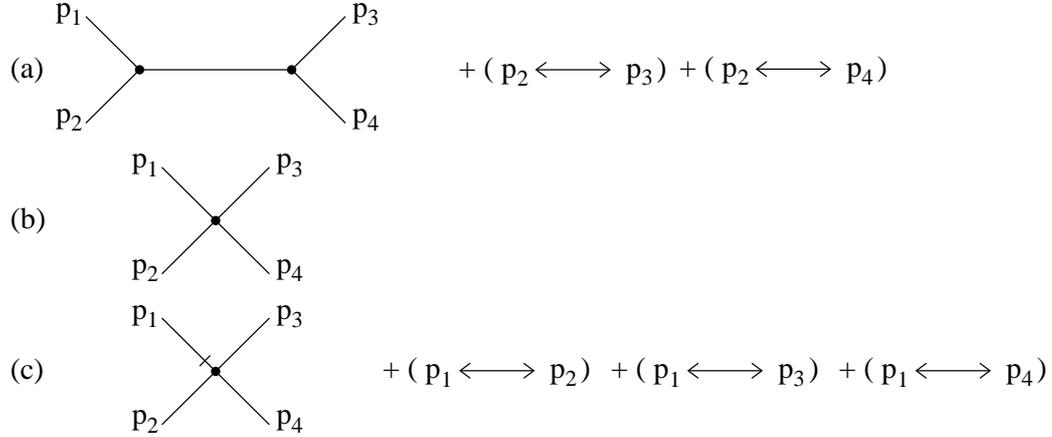} 
\caption{ Various contributions to the 4-point function. (a) Second order
  contributions of the cubic interaction. (b)+(c) Genuine quartic interactions
  without and with a derivative field, the latter being indicated by a
  bar. The various vertex factors are not accounted for.} 
\end{center}
\end{figure}
The most obvious
hint from both equations is that the interactions are concentrated at the 
branch points $ a_i $ , $ i = 1 ,\ldots, 2s $ and that an interaction point is
connected to the external resolvent operator by a 'propagator' $ B_i^0 (p) 
$. \footnote{We will treat the Lagrangean formalism on a formal level without making
concrete use of the fact that the Bergmann kernel $ B(p,q) $ is the Green's 
function of a chiral derivative field on the reduced hyperelliptic
surface as it was noted and used by Dijkgraaf and Vafa \cite{13}.} The 3-point
correlator can obviously be related to a cubic 
interaction:
\begin{equation}
\label{eq: action S_3^0}
S_3^{(0)} = \int \mathcal{L}_3^{(0)} (x) \; dx  \ \ \mbox{  with  }\ \  \mathcal{L}_3^{(0)} (x)
=  \frac{ 1 }{ y_{ 1 } (x) } \frac{ \varphi^3 (x) }{ 3! } = \frac{ 1 }{ y_1 }
\frac{ \varphi^3 }{ 3! }
\end{equation}
with the correspondences
\begin{displaymath}
\int  dx =  \sum_{  i=1 }^{2s}  \Big|_{x=a_i}, \ \ \ \ \ \ \ \ \ \ \ \ \ \ \ \ \ \ \
\ \  
\setlength{\unitlength}{0.250cm}
\begin{picture}(7.2,2)
\put(0.6,2.2){\line(1,0){4.2}}
\put(0.6,2.2){\line(0,-1){1}}
\put(4.8,2.2){\line(0,-1){1}}
\put(0,0){$ \varphi (a_i) \; \varphi (p) $} 
\end{picture} 
= B_i^0 (p) .
\end{displaymath} 

The first bracket on the r.h.s. of (\ref{eq: four-point function}) represents then the second order
contribution of the cubic interaction to the 4-point correlator where one 
propagator is exchanged between the interaction vertices at places $ a_i $ and
$ a_j $. The second and third term on the r.h.s. of (\ref{eq: four-point
  function}) are generated by quartic interactions as 
\begin{equation}
\label{eq: action part a}
 S_4^{(0)(a)} = \int dx \left( - \frac{ 3 y_{2} }{ y_1^3
  } \right) \frac{ \varphi^4 }{4!} 
\equiv \sum_{i=1}^{2s} \left( - \frac{ 3 y_{2,i} }{ y_{1,i}^3 } \right) \frac{ \varphi^4 (a_i) }{4!}
\end{equation}
and
\begin{equation}
\label{eq: action part b}
S_4^{(0)(b)} = \int dx \frac{ 1 }{ y_{1}^2 (x)} \frac{ \varphi^3}{3!} \frac{ \partial \varphi }{ 1 ! }
\equiv \sum_{i=1}^{2s} \frac{ 1 }{ y_{1,i}^2 }  \frac{ \varphi^3 (a_i) }{3!} \frac{ \partial
    \varphi (a_i)}{ 1 ! }
\end{equation}
where the derivative $ (\partial \varphi)_i = (\partial \varphi) (a_i) $ gives rise to the factor $ B_i^1
(p_0 ) $. 
The Lagrange resulting from these quartic interactions is 
\begin{displaymath}
\mathcal{L}_4^{(0)} = \left( - 3 \frac{ y_2 }{ y_1^3 } \right) \frac{
  \varphi^4 }{ 4! } + \left( \frac{ 1 }{ y_1^2 } \right) \frac{ \varphi^3 \;
  \partial \varphi }{ 3!1! }.
\end{displaymath}
We will have to make use of couplings with derivatives of arbitrary
order: $ B_i^f (p) $ will point to an $ f $-th order derivative field at
place $ a_i $, 
\begin{displaymath}
\setlength{\unitlength}{0.250cm}
\begin{picture}(9,2)
\put(2.6,2.5){\line(1,0){3.2}}
\put(2.6,2.5){\line(0,-1){1}}
\put(5.8,2.5){\line(0,-1){1}}
\put(0,0){$ ( \partial^f \varphi )_i \; \varphi (p) $}
\end{picture}
= B_i^f (p)
\ \ \ \ \ \ \ \ \ \ \ \ \ \ 
\setlength{\unitlength}{0.250cm}
\begin{picture}(10,2)
\put(2.6,2.5){\line(1,0){4.9}}
\put(2.6,2.5){\line(0,-1){1}}
\put(7.5,2.5){\line(0,-1){1}}
\put(0,0){$ ( \partial^f \varphi )_i \; ( \partial^g \varphi )_j $}
\end{picture}
= B_{i,j}^{f,g} .
\end{displaymath}
Consecutive application of loop operators  to $ W_4^{(0)} (p_1,p_2,p_3,p_4) $ 
will generate the (connected) correlators of an increasing number of
points. The application of the pieces $ \Delta_4 $ (\ref{eq: B_i^f variation}b) and $ \Delta_7
$ (\ref{eq: B_ij^fg variation}c) lead to the creation of new vertices inserted in external and internal
propagator lines resp., whereas the application of the remaining parts of $
\frac{ \partial }{ \partial V } $ gives rise to a diversification of the
already existing vertices. Let us concentrate for a moment on those contributions to the $ n
$-point function emerging from a single vertex, that is the part of the 
amplitude which will be related to the $ n $-field terms of the effective
Lagrange. There will be terms with $ 3 + k , \ k=0,\ldots, n-3 $ propagators
without derivatives out of the total number $ n $ emerging from an interaction 
point---and there will be corresponding pieces in the Lagrange with the same
number of fields without derivatives. Keeping in mind that the contribution to 
the $ n $-point function going along with the $ n $-field vertices has to be
for itself symmetric in the arguments $ p_1, \ldots , p_n $ one may proceed as follows: One first of all generates $ k $ 
propagators without derivatives by applying $ k $ times $ \Delta_1 $ to the $
3 $-point function, eq. (\ref{eq: three-point function}), and afterwards $
\Delta_2 $ for another $ n -  
(3+k) $ times to generate the propagators with derivatives. The symmetrisation
in $ p_1, \ldots ,p_n $ has to be performed afterwards. That is, the $ 1  
$-vertex contribution to the leading order $ n $-point function is given by 
the symmetrisation of the following expression  
\begin{displaymath}  
\sum_{k=0}^{n-3} \ \overset{n-3-k}{\underset{r=1}{\prod}} \Delta_2 (p_{3+k+r}) 
\overset{k}{\underset{t=1}{\prod}}    \Delta_1  
(p_{3+t} ) \ \  W_3^{(0)} (p_1,p_2,p_3) .  
\end{displaymath} 
One easily extracts therefrom the complete effective Lagrange beyond the two 
lowest orders (eqs. (\ref{eq: action S_3^0})-(\ref{eq: action part b})) 
\begin{equation} 
\label{eq: action}
S^{(0)} = \sum_{n=3}^\infty S_n = \sum_{n \ge 3} \sum_{i=1}^{2s} \mathcal{L}_n^{(0)} ( a_i ) \equiv  
\sum_{i=1}^{2s} \mathcal{L}^{(0)} ( a_i )
\end{equation}  
as a formal infinite series. For the sake of illustration we quote the 
concrete expressions for $ \mathcal{L}_5^{(0)} $ and $ \mathcal{L}_6^{(0)} $  
\begin{equation} 
\mathcal{L}_5^{(0)} = \left( 27 \frac{ y_2^2 }{ y_1^5 } - 15 \frac{ y_3 }{ 
    y_1^4 } \right) \frac{ \varphi^5 }{ 5! } + \left( -9 \frac{ y_2 }{ y_1^4 }  
\right) \frac{ \varphi^4 \; \partial \varphi }{ 4! 1! } + \left( \frac{ 2 }{
    y_1^3 } \right) \frac{ \varphi^3 \; ( \partial \varphi )^2 }{ 3! 2! } +  
\left( 3 \frac{ 1 }{ y_1^3 } \right) \frac{ \varphi^4 ( \partial^2 \varphi
  ) }{ 4! 1! }    
\end{equation} 
\begin{eqnarray} 
\mathcal{L}_6^{(0)} &=& \left( -405 \frac{ y_2^3 }{ y_1^7 } + 450 \frac{ y_2
    y_3 
  }{ y_1^6 } - 105 \frac{ y_4 }{ y_1^5 } \right) \frac{ \varphi^6 }{ 6! }  
+ \left( 135 \frac{ y_2^2 }{ y_1^6 } - 60 \frac{ y_3 }{ y_1^5 } \right) \frac{
  \varphi^5 \; \partial \varphi }{ 5! 1! }
\nonumber\\
&& + \left( - 36 \frac{ y_2 }{ y_1^5 } \right) \frac{ \varphi^4 ( \partial 
  \varphi )^2 }{ 4! 2! } 
+ \left( -54 \frac{ y_2 }{ y_1^5 } \right) \frac{ 
  \varphi^5 ( \partial^2 \varphi ) }{ 5! 1! } 
+ \left( 6 \frac{ 1 }{ y_1^4 } \right) \frac{ \varphi^3 ( \partial \varphi )^3 
}{ 3 ! 3 ! }
\nonumber\\
&& + \left( 9 \frac{ 1 }{ y_1^4 } \right) \frac{ \varphi^4 ( \partial \varphi
  ) ( 
    \partial^2 \varphi ) }{ 4! 1! 1! }
+ \left( 15 \frac{ 1 }{ y_1^4 } \right) \frac{ \varphi^5 ( \partial^3 \varphi 
  ) }{ 5! 1! }  .
\end{eqnarray} \\ \\

We state our first main result as \\ 

{\bf \underline{ Theorem 1: }} \\The connected correlation functions of the   
resolvent operators are in leading order of $ \frac{1}{N} $ given by the tree
graphs of the Gell-Mann-Low series of the effective Lagrange $ 
\mathcal{L}^{(0)} 
$, eq. (\ref{eq: action}). For the determination of an $ n $-point correlator
one has to  
evaluate $ \mathcal{L}^{(0)} $ up to the $ n $-th order, $ \mathcal{L}^{(0,n)} =
\sum_{k=3}^n \mathcal{L}_k^{(0)} $, and to specify all $ n $-point tree graphs 
deducible from $ \mathcal{L}^{(0,n)} $.\\ \\ 
 
The proof of the theorem is a proof by induction. Suppose 
that the statement of the theorem has been found to be true for correlators 
with up to $ n $ points. Applying to the $ n $-point function a further time  
the loop operator one generates either new triple vertices inserted in all 
possible internal and external lines by the action of $ \Delta_7 $  
(eq. (\ref{eq: B_ij^fg variation}c)) and $ \Delta_4 $ (eq. (\ref{eq: B_i^f variation}b)) or one adds a new external line to any of 
the already existing vertices by the action of $ \Delta_1 $, $ \Delta_2 $, $
\Delta_3 $, $ \Delta_5 $ and $ \Delta_6 $. \\Some thought reveals that in this  
way all connected tree graphs with $ (n+1) $ end points related to 
$ \mathcal{L}^{(0,n+1)} \equiv \sum_{k=3}^{n+1} \mathcal{L}^{(0)}_k $  are
generated if one takes the induction assumption for granted.\footnote{One should note
  in this context that the factorials of the Gell-Mann-Low series are
  completely absorbed, as there are no symmetry factors left in tree graph order.}

\section{ Loop corrections } 

\subsection{ 1 loop order}
 
Starting point is the 1-loop version of eq. (\ref{eq: loop eq. to order
h}):

\begin{equation}
\label{eq: loop eq. for h=1}
 W_1^{(1)} (p) = \sum_{ i = 1 }^{2s} \underset{ x \rightarrow a_i }{ \rm Res }
 \ \frac{
      dS (p, x) }{dp} \frac{1}{y(x)} \ W_2^{(0)} (x,x)   .
\end{equation} 
$ dS (p,x) $ and its derivatives can be expressed at the branch points $ x =   
a_r $ by $ B_i^k (p) $ as follows: We define 
\begin{eqnarray*}
 dS (p,[x]_i
)  &=& 2 \; dS( p,x ) \frac{ 1 }{ \sqrt{x-a_i} }
\\
 \widetilde{B}  
  ( p, [x]_i )  &=&  2  \; \widetilde{B} (p,x) \sqrt{x-a_i}  
\\
  y( [x]_i ) &=&
  y (x) \frac{1}{ \sqrt{x-a_i} } 
\end{eqnarray*}
and recalling 
  eq. (\ref{eq: relation between B and dS})
one is immediately lead to
\begin{equation} 
\label{eq: relation between B and dS close to
  branch point}
\widetilde{B} (p,[x]_i ) = \frac{ 1 }{ 4 } \frac{ dS (p,[x]_i ) }{dp} + \frac{ 1   
}{ 2 } (x-a_i) \partial_x \frac{ dS (p ,[x]_i ) }{dp} .
\end{equation}
Differentiation of the last equation with respect to x and then putting $ x =   
a_i $ gives
\begin{equation} 
\label{eq: derivative of dS and B_i^k}
 \partial_x^{\ k}  \big|_{x=a_i} \frac{ dS (p,[x]_i ) }{dp} = \frac{ 4 k! }{ 2k+1 }  
B_i^{k} (p) .
\end{equation} 
From the inspection of the concrete appearance of $ W_2 (p,p) $---that is, the   
form of the Bergmann kernel, of eqs. (\ref{eq: explicit expression for dS
}), (\ref{eq: relation between B and dS}) resp.---we obtain 
\begin{eqnarray}
W_2^{(0)} (x,x) &=& \frac{1}{16} \sum_{i=1}^{2s} \frac{1}{ (x-a_i)^2 } -
\frac{ 1 }{ 16 } \sum_{ i,j = 1 \atop i \neq j }^{2s} \frac{ 1 }{ x - a_i }
\frac{ 1 }{ x - a_j } + \frac{ 1 }{ 4 } \sum_{i=1}^{2s} \frac{ P_i (x) }{
  x-a_i }
\nonumber\\
 &=& \frac{1}{16} \sum_{i=1}^{2s} \frac{ 1 }{ (x-a_i)^2 } -     
\frac 
{1}{8} \sum_{i,j =1 \atop i \neq j}^{2s} \frac{1}{a_i -a_j} \frac{1}{x-a_i} +    
\frac{1}{4} \sum_{i=1}^{2s} \frac{ P_i (a_i) }{x-a_i } 
\nonumber\\ 
 \label{eq: two 
  point function for coinciding arguments}  
&=& \frac{ 1 }{ 16 } \sum_{i=1}^{2s} \frac{1}{(x-a_i)^2} + \frac{ 1 }{ 4 }  
\sum_{i=1}^{2s}\frac{ B_{i,i}^{0,0} }{x-a_i} 
\end{eqnarray}  
where the asymptotic relation $ W_2^{(0)} (p,p) |_{p
  \rightarrow \infty} \sim 1 / p^2 $ is used for the second equality and the last equality can be inferred from a straighforward evaluation of 
the double integral constituting $ B_{i,i}^{0,0} $. The polynomials $ P_m $ 
are related to the polynomials $ L_j $:
\begin{displaymath}
P_m (x) = - \sum_{j=1}^{s-1} L_j (x) \oint\limits_{A_j} \frac{ dz }{ ( z - a_m )
  \sqrt{\sigma(z)} }      \ \ \ \ \ \ \ \ \ \ \ m = 1, \ldots, 2s.
\end{displaymath}

\begin{figure}[h] 
\begin{center}
\epsfig{figure=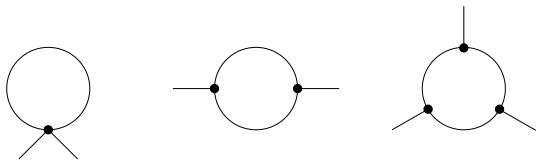, width=7cm}
\caption{1-loop corrections to the 2- and 3-point correlation functions} 
\label{fig: one-loop corrections to 2- and 3-point functions}
\end{center}
\end{figure}

Inserting (\ref{eq: two  
  point function for coinciding arguments}) into
(\ref{eq: loop eq. for h=1})
and taking into account (\ref{eq: derivative of dS and B_i^k}) we arrive at 
\begin{equation}  
\label{eq: one-point function for h=1}
W_1^{(1)} (p) = \sum_{i=1}^{2s} \left( \frac{ 1 }{ 24 } \frac{ 1 }{ y_{1,i} }  
  B_i^1 (p) - \frac{ 1 }{ 8 } \frac{ y_{2,i} }{ ( y_{1,i} )^2 } B_i^0 (p) +
  \frac {1}{2} 
  \frac{ 1 }{ y_{1,i }} B_{i,i}^{0,0} B_i^0 (p) \right) .
\end{equation}
The first two terms on the r.h.s. of (\ref{eq: one-point function for h=1}) emerge from the double pole part in 
eq. (\ref{eq: two
  point function for coinciding arguments}), whereas the third term is due to the part in (\ref{eq: two
  point function for coinciding arguments}) with a single
pole. 
\begin{figure}[h] 
\begin{center}
\epsfig{figure=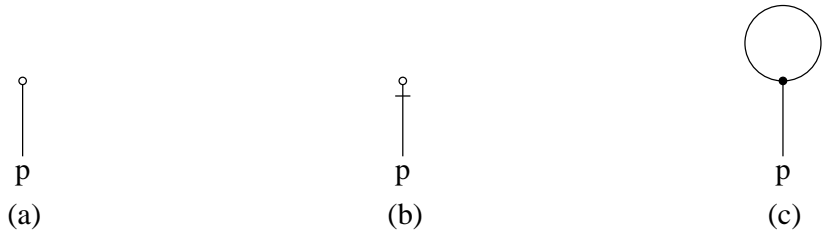} 
\caption{Graphical representation of the $ 3 $ terms contributing to $ W_1^{(1)}
$. The small circles represent vertex insertions with
topological index 1.}
\label{fig: W_1^1} 
\end{center}
\end{figure}
The latter contribution represents the piece which could have been 
anticipated---taking the Lagrangean point of view seriously---as a tadpole
correction of the cubic interaction $ \mathcal{L}_3^{(0)} = \frac{1}{y_1} \frac{ 
  \varphi^3 }{3!} $ to the 1-point function which is graphically depicted in fig. \ref{fig: W_1^1}(c).
The repeated action of the loop operator on the 
tadpole contribution gives rise to the complete set of 1-loop graphs of $
\mathcal{L}^{(0)} = \sum_{n \geq 3} \mathcal{L}_n^{(0)} $ (eq. (\ref{eq: action})), some of those being depicted in 
fig. \ref{fig: one-loop corrections to 2- and 3-point functions}.

The first two terms on the r.h.s. of eq. (\ref{eq: one-point function for h=1}) may be viewed as 'short 
distance corrections' to the above tadpole contribution. To characterize those
terms and their generalisation to be introduced below we ascribe (in a 
slightly ad hoc fashion) to the quantities $ B_i^x $, and $ y_{1+x^{\prime},i}
$ the mass dimensions $ x + 1/2 $ and $ x^{\prime} $ resp., which amounts for
the former quantity to the attachment of a mass dimension $ x + 1/2 $ to a derivative field 
$ \partial^x \varphi $. We also introduce the notion of a topological index $ h 
(V_n) 
$ for an interaction vertex $ V_n $ of $ n $ emanating propagators   
\begin{equation}
 h (V_n) 
= \frac{1}{3} {\rm deg} (V_n) - \frac{1}{2} n + 1 
\end{equation} 
(and the analogous notion for the corresponding part of the interaction  
Lagrange) where the mass dimension $ {\rm deg } (V_n) $ counts the sum
of indices related to ends of propagators emanating from $ V_n $ and the sum
of dimensions of factors $ y_{.\;,\,.} $ attached to $ V_n $. The topological  
index of vertices of the tree graphs considered in the preceding section and
therefore also the Langrange $ \mathcal{L}^{(0)} $ (eq. (\ref{eq: action})) is according to this  
definition zero. The same holds for the vertices of the 1-loop graphs 
displayed in figs \ref{fig: W_1^1}(c) and \ref{fig: one-loop corrections to 2-
  and 3-point functions}. The two short distance corrections in (\ref{eq: one-point function for h=1}) on the   
other hand are of topological index 1, as well as the first non-leading 
correction to the tree graph Lagrange which is obtained by application 
of $ \Delta_1, \Delta_2, \Delta_3, \Delta_5 $ and 
$ \Delta_6 $ (cf. eqs. (\ref{eq: y_fi variation}-\ref{eq: B_ij^fg variation}))
to those terms in question. One finds in this way in particular
\begin{eqnarray}
\label{eq: Lagrange of order 1}
\mathcal{L}^{(1)} &=& \sum_{k=1}^\infty \mathcal{L}_k^{(1)}
\\  
\mathcal{L}_1^{(1)} &=& \left( - \frac{ 1 }{ 8 } \frac{ y_2 }{ y_1^2 } \right) \frac{ \varphi }{
  1! } + \left( \frac{ 1 }{ 24 } \frac{ 1 }{ y_1 } \right) \frac{ \partial
  \varphi }{ 1! }  
\nonumber\\
\mathcal{L}_2^{(1)} &=& \left( \frac{ 3 }{ 4 } \frac{ y_2^2 }{ y_1^4 } - \frac{
    5 }{ 8 } \frac{ y_3 }{ y_1^3 } \right) \frac{ \varphi^2 }{ 2! } 
+ \left( - \frac{ 1 }{ 4 } \frac{ y_2 }{ y_1^3 } \right) \frac{ \varphi \; (
  \partial \varphi ) }{ 1 ! 1 ! }
\nonumber\\
&&+ \left( \frac{ 1 }{ 24 } \frac{ 1 }{ y_1^2 } \right) \frac{ ( \partial
  \varphi )^2 }{ 2! } 
+ \left( \frac{1 }{  8} \frac{1}{y_1^2} \right) \frac{ \varphi \; ( \partial^2
  \varphi ) }{ 1! 1! }
\nonumber\\
\mathcal{L}_3^{(1)} &=& \left( - 9 \frac{ y_2^3 }{y_1^6 } + \frac{ 105 }{ 8 } 
  \frac{ y_2 y_3 }{ y_1^5 } - \frac{ 35 }{ 8 } \frac{ y_4 }{ y_1^4 } \right)
\frac{ \varphi^3 }{ 3! } 
 +\left( 3 \frac{ y_2^2 }{ y_1^5 } - \frac{ 15 }{ 8 } \frac{ y_3 }{ y_1^4 } \right) \frac{ \varphi^2 \;\partial \varphi
}{ 2! 1! }
\nonumber\\
&&+ \left( - \frac{ 3 }{ 4 } \frac{ y_2 }{ y_1^4 } \right) \frac{ \varphi \; ( \partial
  \varphi )^2 }{ 1! 2! }
+ \left( - \frac{ 3 }{ 2 } \frac{ y_2 }{ y_1^4 } \right) \frac{ \varphi^2 \; (
  \partial^2 \varphi ) }{ 2! 1! }
+ \left( \frac{ 1 }{ 12 } \frac{ 1 }{ y_1^3 } \right) \frac{ ( \partial 
  \varphi )^3 }{ 3! } 
\nonumber\\
&&+ \left( \frac{ 1 }{ 4 } \frac{ 1 }{ y_1^3 } \right) \frac{ \varphi \; ( \partial
  \varphi ) \; ( \partial^2 \varphi ) }{ 1! 1! 1! }
+ \left( \frac{ 5 }{ 8 } \frac{ 1 }{ y_1^3 } \right) \frac{ \varphi^2 (
  \partial^3 \varphi ) }{ 2 ! 1! }.
\nonumber
\end{eqnarray}

 To summarize the preceding: The next to leading order contributions to the
 connected correlation functions are given by expressions corresponding to
 \linebreak[2] 1-loop graphs with vertices of topological index zero or to tree graphs with 
 one vertex from eq. (\ref{eq: Lagrange of order 1}) with topological index 1.

\subsection{  More loops }  
\subsubsection{  A preparatory calculation }
Proceeding from order $ (h-1) $ to $ h $ in the topological expansion one has  
to resort to the loop equation (\ref{eq: loop eq. to order
h}). An immediate consequence of the recursive nature of  equation (\ref{eq: loop eq. to order
h}) is that the 1- and 2-point functions have the general form 
\begin{equation} 
\label{eq: separation of outer legs in W_1^h}
W_1^{(h)} (x) = \sum_{i,f} B_i^f (x) \omega_1^{(h)} (i,f)   
\end{equation}
and  
\begin{equation} 
W_2^{(h)} (x,x) = \sum _{ i,j \atop f,g } B_i^f (x) B_j^g (x) 
\omega_2^{(h)} (i,f;j,g) 
\end{equation}
where the functions $ \omega_k^{(h)} $ do not depend on $ x $.
Confronting eq. (\ref{eq: loop eq. to order
h})  with the last two equations we see that we have to deal 
with the evaluation of expressions
\begin{displaymath}
\underset{ x \rightarrow a_i }{\rm Res } \frac{ dS (p,x) }{dp} \frac{ 1 }{
  y(x) } B_j^f (x) B_k^g (x)  .
\end{displaymath}
We do this for arbitrary integer values $ f $ and $ g $ and the special (most  
involved) case $ i = j = k $, merely quoting at the end the results for the
other (simpler) cases, which are practically contained in the case of  
coincinding arguments. Cauchy's integral representation for the residuum and
the representation eq. (27a) for $ B^f_i $ and $ B^g_i $ lead to   
\begin{eqnarray}
&\underset{ x \rightarrow a_i }{\rm Res}& \frac{ dS (p,x) }{dp } \frac{1}{y(x)}   
B_i^f (x) B_i^g (x)  
\nonumber\\
\label{eq: integral
  contour twist step one}
&=& \oint\limits_{a_i} \frac{dx}{2 \pi i} \frac{dS (p,x) }{dp \ 
  y(x)} \ 2 \oint\limits_{a_i} \frac{ dw }{ 2 \pi i } \frac{ \widetilde{B} (w,x) 
}{ ( w - a_i )^{f+1/2} } \ 2 \oint\limits_{a_i} \frac{ dv }{ 2 \pi i } \frac{ \widetilde{B}
  (v,x) }{ 
  ( v - a_i )^{g+1/2} } .\ \  
\end{eqnarray}
The contour in $ x $ is by definition positioned outside the contours of those 
for the variables $ w $ and $ v $. The order of the latter contours is
immaterial but we assume for the sake of definiteness that the contour in $ w 
$ surrounds that in $ v $. Pushing the $ x $-contours in (\ref{eq: integral
  contour twist step one}) through the
other two one picks up two additional terms from the double poles of $ 
\widetilde{B} (w,x) $ and $ \widetilde{B} (v,x) $ at $ w=x $ and $ v = x $ resp.,  
\begin{eqnarray}
(\ref{eq: integral 
  contour twist step one}) = 4 \oint\limits_{a_i} \frac{ dw }{ 2 \pi i } 
\frac{ 1 }{ (w - a_i)^{f+1/2} }   
 \oint\limits_{a_i} \frac{ dv }{ 2 \pi i } \frac{ 1 }{ ( v - a_i )^{g+1/2} }
\oint\limits_{a_i} \frac{ dx }{ 2 \pi i } \frac{ dS (p,x) }{ dp \; y(x) }
 \widetilde{B}
(w,x)  
\widetilde{B} (v, x)   
\nonumber\\
{} + 4 \oint\limits_{a_i} \frac{ dw }{ 2 \pi i } \frac{ 1 }{ (w - a_i )^{f+1/2} } \oint\limits_{a_i} \frac{ dv }{ 
  2 \pi i } \frac{ 1 }{ ( v - a_i )^{g+1/2} } \left( \frac{ 1 }{ 2 }
  \partial_w 
  \left( \frac{ dS (p,w) }{ dp \; y (w) } \widetilde{B} (w,v) \right) \right.
\nonumber\\
\label{eq: integral 
  contour twist step two}
\left. {} + \frac{ 1 }{ 2 } \partial_v  
  \left( \frac{ dS (p,v) }{ dp \; y (v) } \widetilde{B} (v,w) \right)  \right)
. \ 
\end{eqnarray}
The first term on the r.h.s., let us denote it as (\ref{eq: integral
  contour twist step two}a), where the $ x  
$-contour is now the most inner one, can easily be computed since the
integrand has a simple pole at $ x = a_i $. Taking into account eq. (\ref{eq: derivative of dS and B_i^k})  we 
obtain 
\begin{displaymath} 
{\rm (\ref{eq: integral
  contour twist step two}a) } = \frac{1}{2} \frac{ B_i^0 (p) B_{i,i}^{f,0}
B_{i,i}^{g,0} }{   y_{1,i} } . 
\end{displaymath}
The two other terms on the r.h.s. of (\ref{eq: integral
  contour twist step two}), denoted (\ref{eq: integral
  contour twist step two}b) and (\ref{eq: integral
  contour twist step two}c) resp., 
become after partial integration
\begin{equation}
\label{eq: integral contour twist step three} 
{\rm (\ref{eq: integral 
  contour twist step two}b)} = ( 2f+1 ) \oint\limits_{a_i} \frac{ dw }{ 2 \pi i } \frac{ 1
}{ (w-a_i)^{f+3/2} 
} 
  \oint\limits_{a_i} \frac{ dv }{ 2 \pi i } \frac{ 1 }{ ( v - a_i )^{g+1/2} } \frac{ dS (p ,w  )}{dp } \frac{ \widetilde{B} ( w, v ) }{ y (w) } 
\end{equation}
and 
\begin{equation}
\label{eq: integal contour twist step four} 
{\rm (\ref{eq: integral
  contour twist step two}c)} = ( 2g+1 ) \oint\limits_{a_i} \frac{ dw }{ 2 \pi i } \frac{ 1
}{ (w-a_i)^{f+1/2} }  
  \oint\limits_{a_i} \frac{ dv }{ 2 \pi i } \frac{ 1 }{ ( v - a_i )^{g+3/2} } \frac{ dS (p ,v   )}{dp } \frac{ \widetilde{B} ( v, w ) }{ y (v) }.
\end{equation}
The $ v $-integration in the last equation---being now the most inner 
one---can immediately be executed and one finds with eq. (\ref{eq: derivative of dS and B_i^k})
\begin{equation} 
\label{eq: integal contour twist step five}
{\rm (\ref{eq: integal contour twist step four})} = (g+\frac{1}{2})  \sum_{ k
  + r + t = g + 1 } \frac{ 1 }{ 2 k +1 }
B_i^k    (p) B_{i,i}^{f,r} Z_{t,i} \mbox{ with} Z_{t,i} = \frac{ \partial_x^{ \ t} }{ t! }
\frac{ 
  1 }{ y ([x]_i ) } \bigg|_{x=a_i} .
\end{equation}
The evaluation of (\ref{eq: integral contour twist step three}) proceeds by pushing the $ w $ contour through the $ v
$-contour. One obtains in this way one term analogously to eq. (\ref{eq:
  integal contour twist step five}) and an additional term from the pole at $
w = v $ yielding:
\begin{eqnarray}
\label{eq: integal contour twist step six}
{\rm (\ref{eq: integral contour twist step three})} = (f+\frac{1}{2}) \sum_{k+r+t=f+1} \frac{ 1 }{ 2 k +1 } B_i^k (p)
 B_{i,i}^{g,r} Z_{t,i}
\nonumber\\
+ \frac{ 1 }{ 2 } (2 f + 1 )(2 g + 1  ) \sum_{ k+ l =f+g+2} \frac{1}{2k+1}
B_i^k (p) Z_{l,i} .
\end{eqnarray}
Putting pieces together, i.e. eqs. (\ref{eq: integral
  contour twist step two}a), (\ref{eq: integal contour twist step five}) and (\ref{eq: integal contour twist step six}) and taking care of the
cases of non-coincident indices, one arrives at
\begin{eqnarray}
&\underset{ x \rightarrow a_i }{\rm Res }& \frac{ dS (p,x) }{dp\; y(x) } B_j^f 
(x) B_k^g (x)  
\nonumber\\
&=&  \frac{1}{2} B_{j,i}^{f,0} \frac{ B_i^0 (p) }{ y_{1,i} } 
B_{i,k}^{0,g} 
\nonumber\\
&+& \delta_{k,i} (g+\frac{1}{2}) \sum_{ r + m + t = g + 1 } \frac{ 1 }{ 2 r + 1 } 
B_i^r (p)  B_{j,i}^{f,m} Z_{t,i}
\nonumber\\ 
&+& \delta_{j,i} (f+\frac{1}{2}) \sum_{ r + m + t = f + 1 } \frac{ 1 }{ 2 r + 1 }
B_i^r (p)  B_{k,i}^{g,m} Z_{t,i}
\nonumber\\
\label{eq: integral contour twist result}
&+& \delta_{k,i} \delta_{j,i} \frac{1}{2} ( 2 g+1 ) (2 f + 1) \sum_{ r + l = f +
  g +2 } \frac{1 }{2r+1} B_i^r (p) Z_{l,i} .
\end{eqnarray}
It is worth noting for later reference that from all terms on the r.h.s. of  
eq. (\ref{eq: integral contour twist result}) only the last term is responsible for the 
creation of a vertex with an increased topological index (The index of the
fused vertex is then equal to the sum of the two (non-fused) original vertices
if it results from the first term on the r.h.s. of eq. (\ref{eq: loop eq. to order
h}) or it increases by one unit if it originates from the second term.)

\subsubsection{ Free energy at higher loop orders }
We adapt momentarily our notations to those of ref. \cite{5} by writing $ Z(t) = \int 
dM \exp (-\frac{N}{t_0} {\rm tr} V (M)) = e^{\mathcal{F}} $ with $ V (M)=
\sum_{n \geq 
  1 }  t_n  M^n $ and remind that the occupation numbers were above 
designated by $ \varepsilon_i $. Chekhov and Eynard \cite{5} noted that the 
scaling relation for the free energy 
\begin{displaymath} 
\left( \sum t_k \frac{\partial }{ \partial t_k } + t_0 \frac{ \partial }{ 
    \partial t_0 } + \sum \varepsilon_i \frac{\partial}{\partial \varepsilon_i}
    + h \frac{ \partial }{ \partial h} \right) \mathcal{F} = 0 , 
\end{displaymath} 
with $ h = \frac{ t_0 }{ N} $, is synonymous with a formula which gives $ 
\mathcal{F}^{(k)} $  as 'integral' of $ W_1^{(k)} $. The details of the integration  
operator, called $ H $ in \cite{5}, is immaterial for our purposes. We have  
only to note
that $ H $ applied to $ B $ gives  
\begin{displaymath}
H \cdot \frac{ B(\cdot,p) }{ dp } = - \frac{1}{2} y (p) 
\end{displaymath} 
to find with the help of (\ref{eq: separation of outer legs in W_1^h}) and the
previous scaling relation  
\begin{eqnarray}
\label{eq: free energy by H operator}
\mathcal{F}^{(h)} = - \frac{1}{2 h- 2} H_q \cdot W_1^{(h)} (q)   
=  \frac{ 1}{2 h-2} \sum_{i \atop l \geq 1 }  
\omega_1^{(h)} (i,l) y_{l,i} 
\end{eqnarray}
where the $ l = 0 $ part of eq. (\ref{eq: separation of outer legs in W_1^h}) drops out since it leads to a vanishing 
residuum. Eq. (\ref{eq: free energy by H operator}) holds for $ h \geq 2 $. What concerns the expressions for $
\mathcal{F}^{(1)} $ and $ \mathcal{F}^{(0)} $ we refer to \cite{5} and
\cite{8} and in particular \cite{21}. 

\subsubsection{  Short distance corrections }

We want to work out the repercussions of the last term in (\ref{eq: integral contour twist result}) on the higher order corrections to the free
energy, the $ 1 $-point function and the effective Lagrange resp.. We 
restrict here our attention to 'local' contributions, i. e. to those
quantities which only depend on functions $ y_{x,i} $ and $ B_i^x $ from  
one and the same branch point at a time and do in particular not depend on  
propagators $ B_{i,i}^{x,y} $, $ B_{i,j}^{x^\prime, y^\prime} $ connecting the  
same or different branch points. We use here and in the following a hat to designate 'local' quantities. Starting from the short distance corrections
to $ W_1^{(1)} (p) $, the two first terms on the r.h.s. of eq. (\ref{eq:
  one-point function for h=1}),    
\begin{displaymath}
\ \ \ \ \ \ \ \ \ \ \ \ \ \ \ \ \ \ \ \ \ \ \ \ \ \ \widehat{W}_1^{(1)} (p) = \sum_i \left( \frac{ 1}{24} \frac{1}{ y_{1,i}}  B_i^1 
  (p) - 
  \frac{1}{8} \frac{ y_{2,i} }{ y_{1,i}^3 } B_i^0 (p) \right) ,
\ \ \ \ \ \ \ \ \ \ \ \ \ \ \ \ \ \ \ \ \ \ \ \ \ \ (\ref{eq: one-point function for h=1}a)
\end{displaymath}
one obtains a 'local' contribution to $ \widehat{W}_2^{(1)} (p,q) $ by applying 
the parts $ \Delta_1, \Delta_2 $ and $ \Delta_3 $ of $ \frac{ \partial }{
  \partial V} $ to $ \widehat{W}_1^{(1)} $:
\begin{eqnarray} 
\widehat{ W }_2^{(1)} (p_1 , p_2 ) &=& 
\sum_{i=1}^{2s} \left( \frac{ 3 }{ 4 } \frac{ y_2^2 }{ y_1^4 } - \frac{ 5 }{ 8
  } \frac{ y_3 }{ y_1^3 } \right)_i B_i^0 (p_1 ) \; B_i^0 (p_2)
\nonumber\\
&&+ \sum_{i=1}^{2s} \left( - \frac{ 1 }{ 4 } \frac{ y_2 }{y_1^3} \right)_i
\left( B_i^1 ( p_1 ) B_i^0 (p_2) + B_i^0 (p_1) B_i^1 (p_2) \right)
\nonumber\\
&&+ \sum_{i=1}^{2s} \left( \frac{ 1 }{ 8 } \frac{ 1 }{ y_1^2 } \right)_i \left(
  B_i^2 (p_1) B_i^0 (p_2) + B_i^0 (p_1) B_i^2 (p_2) \right)
\nonumber\\
\label{eq: hat_W_2^1}
&&+ \sum_{i=1}^{2s} \left( \frac{ 1 }{ 24 } \frac{ 1 }{ y_1^2 } \right)_i B_i^1
(p_1) \; B_i^1 (p_2) .
\end{eqnarray} 
Inserting the local part of eq. (\ref{eq: one-point function for h=1}) and (\ref{eq: hat_W_2^1}) into the loop eq. (\ref{eq: loop eq. to order
h}) and restricting to the last part of eq. (\ref{eq: integral 
  contour twist result}) one finds  
\begin{eqnarray}  
\widehat{W}_1^{(2)} (p_1) &=&
 \sum_{i=1}^{2s} \left(  \frac{ 63 }{ 32 } \frac{ y_2^4 }{ y_1^7 } - \frac{ 75  
   }{ 16 } \frac{ y_2^2 y_3 }{ y_1^6 } + \frac{ 77 }{ 32 } \frac{ y_2 y_4 }{ 
     y_1^5 } + \frac{ 145 }{ 128 } \frac{ y_3^2 }{ y_1^5 } - \frac{ 105 }{ 128 
 } \frac{ y_5 }{ y_1^4 } \right)_i B_i^0 (p_1)
\nonumber\\
&&+ \sum_{i=1}^{2s} \left( - \frac{ 21 }{32 } \frac{ y_2^3 }{ y_1^6 } + \frac{  
    29 
  }{ 32 } \frac{ y_2 y_3 }{ y_1^5 } - \frac{ 35 }{ 128 } \frac{ y_4 }{ y_1^4 } 
\right)_i B_i^1 (p_1) 
\nonumber\\
&&+ \sum_{i=1}^{2s} \left( \frac{ 63 }{ 160 } \frac{ y_2^2 }{y_1^5} - \frac{  
    29 }{ 128 } \frac{ y_3 }{ y_1^4 } \right)_i B_i^2 (p_1)
\nonumber\\ 
&&+ \sum_{i=1}^{2s} \left( - \frac{ 29 }{ 128 } \frac{ y_2 }{ y_1^4 }
\right)_i 
B_i^3 (p_1)
\nonumber\\ 
\label{eq: local part of one loop function for h=2}
&&+ \sum_{i=1}^{2s} \left( \frac{ 35 }{ 384 } \frac{ 1 }{ y_1^3 } \right)_i 
B_i^4 (p_1)
\end{eqnarray}  
and from eq. (\ref{eq: local part of one loop function for h=2}) a local
contribution to the free energy  
\begin{equation} 
 \widehat{\mathcal{F}}^{(2)}  = 
 \sum_{i=1}^{2s} \left( - \frac{ 21 }{ 160 } \frac{ y_2^3 }{ y_1^5 } + \frac{ 29
   }{ 128 } \frac{ y_2 y_3 }{ y_1^4 } - \frac{ 35 }{ 384 } \frac{ y_ 4 }{ y_1^3
   } \right)_i .
\end{equation} 
Applying the loop operator to $ \widehat{W}_1^{(2)} (p) $ one finds $  
\widehat{W}_2^{(2)} (p_1,p_2) $, $ \widehat{W}_3^{(2)} (p_1,p_2,p_3) $  etc. and can extract from this the two loop short distance correction to the effective 
Lagrange
\begin{eqnarray}
  \mathcal{L}_2^{(2)} = \left( 
-\frac{ 1323 }{ 32 } \frac{ y_2^5 }{ y_1^9 }   
+\frac{ 495 }{ 4 } \frac{ y_2^3 y_3 }{ y_1^8 } 
-\frac{ 2205 }{ 32 } \frac{ y_2^2 y_4 }{ y_1^7 }  
-\frac{ 8175 }{ 128 } \frac{ y_2 y_3^2 }{ y_1^7 } 
+\frac{ 63 }{ 2 } \frac{ y_2 y_5 }{ y_1^6 }  
+\frac{ 1785 }{ 64 } \frac{ y_3 y_4 }{ y_1^6 } \right.
\nonumber\\
\left. {}
-\frac{ 1155 }{ 128 } \frac{ y_6 }{ y_1^5 }  
\right) \frac{ \varphi^2 }{ 2! }
+ \left(
\frac{ 441 }{ 32 } \frac{ y_2^4 }{ y_1^8 }   
-\frac{ 225 }{ 8 } \frac{ y_2^2 y_3 }{ y_1^7 } 
+\frac{ 385 }{ 32 } \frac{ y_2 y_4 }{ y_1^6 }  
+\frac{ 725 }{ 128 } \frac{ y_3^2 }{ y_1^6 } 
-\frac{ 105 }{ 32 } \frac{ y_5 }{ y_1^5 } 
\right) \frac{ \varphi \; ( \partial \varphi ) }{ 1!1! }
\nonumber\\ 
+ \left(
-\frac{ 63 }{ 8 } \frac{ y_2^3 }{ y_1^7 }   
+\frac{ 75 }{ 8 } \frac{ y_2 y_3 }{ y_1^6 } 
-\frac{ 77 }{ 32 } \frac{ y_4 }{ y_1^5 } 
\right) \frac{ \varphi \; ( \partial^2 \varphi ) }{ 1!1! }
+\left(
-\frac{ 63 }{ 16 } \frac{ y_2^3 }{ y_1^7 }   
+\frac{ 145 }{ 32 } \frac{ y_2 y_3 }{ y_1^6 } 
-\frac{ 35 }{ 32 } \frac{ y_4 }{ y_1^5 } 
\right) \frac{ ( \partial \varphi )^2 }{ 2! }
\nonumber\\
+\left(
\frac{ 75 }{ 16 } \frac{ y_2^2 }{ y_1^6 }   
-\frac{ 145 }{ 64 } \frac{ y_3 }{ y_1^5 } 
\right) \frac{  \varphi \; (\partial^3 \varphi ) }{ 1!1! } 
+\left(
\frac{ 63 }{ 32 } \frac{ y_2^2 }{ y_1^6 }   
-\frac{ 29 }{ 32 } \frac{ y_3 }{ y_1^5 } 
\right) \frac{ ( \partial \varphi ) \; (\partial^2 \varphi ) }{ 1!1! } 
\nonumber\\
+\left(
-\frac{ 77 }{ 32 } \frac{ y_2 }{ y_1^5 }   
\right) \frac{  \varphi  \; (\partial^4 \varphi ) }{ 1!1! }
+\left(
-\frac{ 29 }{ 32 } \frac{ y_2 }{ y_1^5 }   
\right) \frac{  ( \partial \varphi ) \; ( \partial^3 \varphi ) }{ 1!1! }
+\left(
-\frac{ 63 }{ 80 } \frac{ y_2 }{ y_1^5 }   
\right) \frac{  ( \partial^2 \varphi )^2 }{ 2! }
\nonumber\\
+\left(
\frac{ 105 }{ 128 } \frac{ 1 }{ y_1^4 }  
\right) \frac{  \varphi  \; (\partial^5 \varphi ) }{ 1!1! }
+\left(
\frac{ 35 }{ 128 } \frac{ 1 }{ y_1^4 }  
\right) \frac{  ( \partial \varphi )  \; ( \partial^4 \varphi ) }{ 1!1! }
+\left(
-\frac{ 29 }{ 128 } \frac{ 1 }{ y_1^4 }  
\right) \frac{  ( \partial^2 \varphi )  \; (\partial^3 \varphi ) }{ 1!1! } .\ 
\end{eqnarray}
It should by now be clear how to proceed to the determination of higher order    
local quantities. Let us assume that $ \widehat{W}_1^{(1)} (p) , \ldots, 
\widehat{W}_1^{(h)} (p) $ have been determined. One obtains $ \widehat{W}_2^{(h)}  
(p,q) $ by loop differentiation of $ \widehat{W}_1^{(h)} $ and then $  
\widehat{W}_1^{(h+1)} (p) $ by insertion of $ \sum_{m=1}^h \widehat{W}_1^ {(h+1-m)} (x)
\widehat{W}_1^{(m)} (x) + \widehat{W}_2^{(h)} (x,x) $ into the last term of
eq. (\ref{eq: integral contour twist result}) and therefrom $
\widehat{\mathcal{F}}^{(h+1)} $. One arrives in this manner for example at
\begin{eqnarray} 
\widehat{\mathcal{F}}^{(3)}= 
\sum_{i=1}^{2s} \left(
\frac{2205}{256} \frac{y_2^6}{y_1^{10}}
-\frac{8685}{256} \frac{y_2^4 y_3}{y_1^9}
+\frac{15375}{512} \frac{y_2^2 y_3^2}{y_1^8}
+\frac{5565}{256} \frac{y_2^3 y_4}{y_1^8}
-\frac{72875}{21504} \frac{y_3^3}{y_1^7} \right.
\nonumber\\
\left.
-\frac{5605}{256} \frac{y_2 y_3 y_4 }{y_1^7}
-\frac{3213}{256} \frac{y_2^2 y_5}{y_1^7}
+\frac{21245}{9216} \frac{y_4^2}{y_1^6}
+\frac{2515}{512} \frac{y_3 y_5}{y_1^6}
+\frac{5929}{1024} \frac{y_2 y_6}{y_1^6}
-\frac{5005}{3072} \frac{y_7}{y_1^5}
\right)_i .\ 
\end{eqnarray}
In the table of all single Lagrange functions $ \mathcal{L}_k^{(h)} $ two pieces in $ h=0 $ are missing because we start with $ W_3^{(0)} $\footnote{$
  W_1^{(0)} $ and $ W_2^{(0)} $ instead supply the ingredients of the above
  introduced graphs, the vertex factors and the propagators.}. We absorb from 1-loop order onwards the function $ \widehat{W}_1^{(h)} $ as $ \mathcal{L}_1^{(h)} $ into the Lagrange and get finally:
\begin{eqnarray*} 
\mathcal{L}^{(0)} &=& \phantom{\mathcal{L}_1^{(0)}}  \phantom{+}
\phantom{\mathcal{L}_2^{(0)}} \phantom{+} \ \ \ \! \: \mathcal{L}_3^{(0)} + \mathcal{L}_4^{(0)} + \ldots 
\\
\mathcal{L}^{(h)} &=& \mathcal{L}_1^{(h)} + 
\mathcal{L}_2^{(h)} + \mathcal{L}_3^{(h)} + \mathcal{L}_4^{(h)} + \ldots  \ \
\ \ \ \mbox{ for } h \ge 1 .
\end{eqnarray*}
One may also introduce $ \mathcal{L}_0^{(h)} $, $ h \ge 2 $, as the local part
$ \widehat{\mathcal{F}}^{(h)} $ of the free energy.
What concerns the full free energy $ \mathcal{F}^{(h)} $, cf. eq. (\ref{eq: free 
  energy by H operator}). \\
To take the two special cases appearing in the Lagrange also out of the 
correlators, one has to define  
\begin{displaymath}
\overline{W}_k = W_k - \delta_{k,1} W_1^{(0)} - \delta_{k,2} W_2^{(0)} .
\end{displaymath} 
The expression $ \langle 0 | \ldots | 0 \rangle_{l \ \rm loops,\ conn} $ denotes the 
sum of all $ l $ loop diagrams of $ \langle 0 | \ldots | 0 \rangle_{conn} $.\\

\begin{figure}[h] 
\begin{center}
\epsfig{figure=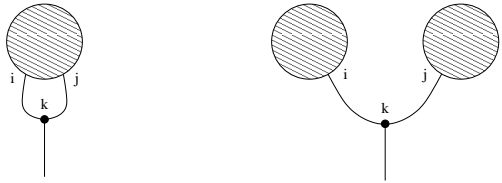,  width=6.5cm } 
\caption{Contributions to $ W_1^{(h)} $  } 
\label{fig: Contributions to W_1^h} 
\end{center} 
\end{figure}

\begin{figure}[h]
\begin{center}
\epsfig{figure=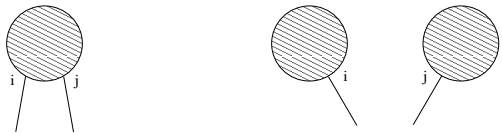,  width=6.5cm }
\caption{Lower order amplitudes related to those of fig. \ref{fig: Contributions to W_1^h}} 
\label{fig: lower order amplitudes}
\end{center}
\end{figure}

The main result of the present paper is comprised in\\ \\
\underline{\bf Theorem 2:} \\
 With the
Lagrange $ \mathcal{L} = \mathcal{L}^{(0)} + \frac{ 1 }{ N^2 } \; \mathcal{L}^{(1)} + \frac{
  1 }{ N^4 } \; \mathcal{L}^{(2)} + \ldots 
$
the Hermitean 1-matrix model correlation functions are given by
\begin{displaymath}
\overline{W}_k ( p_1 ,\ldots , p_k ) =\sum_{l=0}^\infty \frac{ 1 }{ N^{ 2l } } \; 
 \langle 0 | \; \varphi (p_1) \cdots \varphi (p_k)
\ e^{ \sum_{ i=1 }^{2s} \mathcal{L} (a_{i}) } \, | 0 \rangle_{ l \ \rm loops,\
conn} . 
\end{displaymath}
For the evaluation of the Gell-Mann-Low series the propagators \\ \\  
\setlength{\unitlength}{0.250cm}
\begin{picture}(9,2) 
\put(2.6,2.5){\line(1,0){3.2}}
\put(2.6,2.5){\line(0,-1){1}} 
\put(5.8,2.5){\line(0,-1){1}} 
\put(0,0){$ ( \partial^f \varphi )_i \; \varphi (p) $}
\end{picture}
and
\setlength{\unitlength}{0.250cm}
\begin{picture}(10,2)
\put(2.6,2.5){\line(1,0){4.9}}
\put(2.6,2.5){\line(0,-1){1}}
\put(7.5,2.5){\line(0,-1){1}}
\put(0,0){$ ( \partial^f \varphi )_i \; ( \partial^g \varphi )_j $}
\end{picture}
have to be taken as $ B_i^f (p)$ and $  B_{i,j}^{f,g} $ resp.. \\ \\ 
\underline{\bf sketch of proof:} One has first of all to show that the    
interaction vertices are all generated by the repeated action of the loop  
operator on $ \widehat{W}_1^{(h)} $. This can be proved inductively: Assuming that  
it is found true to $ (n-1) $-th order the step to $ n $-th order is done  
through the 
loop equation (\ref{eq: loop eq. to order 
h}). It is in this context 
particularly convenient to consider the free energy $ \mathcal{F}^{(h)} $. All
terms of  
eq. (\ref{eq: integral contour twist result}) besides the last term on the
r.h.s. of eq. (\ref{eq: integral contour twist result}) (and leading to the local 
correction $ \widehat{\mathcal{F}}^{(h)} $) have a graphical appearance as shown in
fig. \ref{fig: graphical appearance of terms contributing to the free energy not being of local type}. 
\begin{figure}[h]
\begin{center}
\epsfig{figure=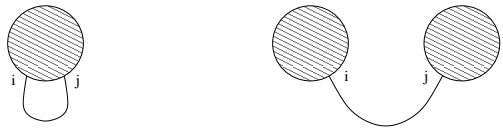, width=6.5cm } 
\caption{Graphical appearance of terms contributing to the free energy $
  \mathcal{F}^{(h)} $ not being of the local type $ \widehat{\mathcal{F}}^{(h)} $}
\label{fig: graphical appearance of terms contributing to the free energy 
  not being of local type}
\end{center} 
\end{figure}
Those contributions to the free energy depicted in fig. \ref{fig: graphical appearance of terms contributing to the free energy
  not being of local type} are related to contributions to $ 
W_1^{(h)}  (p) $ of fig. \ref{fig: Contributions to W_1^h} via action of loop differentiation. 
The amplitudes shown in fig. \ref{fig: Contributions to W_1^h} are uniquely  
connected by the
loop equations to lower order terms given in fig. \ref{fig: lower order amplitudes}. (We select for the purpose of
our argument a term with $ k \neq i,j $ which does not give rise to short  
distance corrections at the extra vertex placed at $ a_k $.) 
But for the latter amplitudes holds the inductive assumption and therefore  
also for all contributions to $ \mathcal{F}^{(h)} $ which are not belonging to the local
part $ \widehat{\mathcal{F}}^{(h)} $ of $ \mathcal{F}^{(h)} $. It is then in particular true that the local vertices  
making up $ \mathcal{F}^{(h)} $ are those which were already present in $ \mathcal{F}^{(h-1)} $
(covered by the inductive assumption) and the new local vertex for $ \widehat{\mathcal{F}}^{(h)} $ (and then also for  
$ \widehat{W}_1^{(h)}, \mathcal{L}_2^{(h)} , $ etc.). This concludes the
argument.\\  
The missing part for a full proof of the above theorem, that is, a control of 
symmetry numbers in 2 loop order\footnote{The proper appearance of the 1-loop order
has been checked.} and beyond, requires a detailed
combinatorical analysis and will be given in a separate forthcoming
publication \cite{19}.
\section{ Summary and conclusions } 
The purpose of the paper was to develop a Lagrangean formalism of Hermitean  
1-matrix models taking as starting point Eynard's technique \cite{4} encoded
in the hyperelliptic Riemann surface associated as spectral surface to the  
matrix model. Eynard's approach is based on 2 ingredients: 1.) The differential 1-form attached to the defining equation of the 
hyperelliptic surface and \linebreak[4] 2.) the two-differential, the Bergmann kernel, on
the reduced surface. Those ingredients are found to show up in the solution of the loop equations for the connected correlation function of resolvent operators
given by an iterative nested system of Cauchy contour integrals. We resolved
this nested system into the rules for an effective Lagrange
of a scalar field propagating on the reduced surface, s.t. the propagation is
by the Bergmann kernel with multiple self-interactions of the scalar field
taking place at the branch points. The interactions are represented as a
formal infinite power series in powers of the scalar fields and its
derivatives and are subject to short distance corrections---due to the
singularity of the Bergmann kernel at coinciding points---at all orders of the
topological expansion. 
Our motivation for the endeavour to construct a Lagrange formalism is to find a new approach to critical
behaviour via the renormalization group. The test of the usefulness of the
formalism in this context is still ahead of us.\\
Evidently, the formalism reveals an abundantly rich true Lagrangean structure behind Eynard's
trivalent graphical representation.\\

\end{document}